# Wide optical spectrum range, sub-volt, compact modulator based on electro-optic polymer refilled silicon slot photonic crystal waveguide

**Xingyu Zhang,[1,*] Amir Hosseini,[2,*] Jingdong Luo,[3] Alex K.-Y. Jen,[3] and Ray T. Chen[1,*]**

[1]*Microelectronics Research Center, Electrical and Computer Engineering Department, University of Texas at Austin, Austin, TX, 78758, USA*
[2]*Omega Optics, Inc., 10306 Sausalito Dr., Austin, TX 78759, USA*
[3]*Department of Materials Science and Engineering, University of Washington, Seattle, Washington 98195, USA*
**Corresponding authors: xzhang@utexas.edu, amirh@utexas.edu, raychen@uts.cc.utexas.edu*

We design and demonstrate a compact and low-power band-engineered electro-optic (EO) polymer refilled silicon slot photonic crystal waveguide (PCW) modulator. The EO polymer is engineered for large EO activity and near-infrared transparency. A PCW step coupler is used for optimum coupling to the slow-light mode of the band-engineered PCW. The half-wave switching-voltage is measured to be $V_\pi$=0.97±0.02V over optical spectrum range of 8nm, corresponding to the effective in-device $r_{33}$ of 1190pm/V and $V_\pi \times L$ of 0.291±0.006V×mm in a push-pull configuration. Excluding the slow-light effect, we estimate the EO polymer is poled with an efficiency of 89pm/V in the slot.

Electro-optic (EO) polymer modulators in optical links are promising for low power consumption [1] and broad bandwidth operation [2]. The electro-optic coefficient ($r_{33}$) of EO polymers can be several times larger than that of lithium niobate. In addition to conventional all-polymer devices [1, 2] , combination of silicon photonics and EO polymer have shown to enable compact and high performance integrated devices [3], such as slot waveguide Mach-Zehnder Interferometer (MZI) modulators [4], slot waveguide ring-resonator modulators [5], and slot Photonic Crystal Waveguide (PCW) modulators [6]. The fabrication process of these devices involves the poling of the EO polymer at an elevated temperature. Unfortunately, the leakage current due to the charge injection through silicon/polymer interface significantly reduces the poling efficiency in narrow slot waveguides (slot width, $S_w$<200nm). Among the abovementioned structure, the slot PCW can support optical mode for $S_w$ as large as 320nm [7]. Such a wide slot was shown to reduce the leakage current by two orders of magnitude resulting in 5x improvement in the in-device $r_{33}$ compared to a slot PCW with $S_w$=75nm [7].

One problem remains among slot PCW modulators is their narrow operating optical bandwidth of <1nm [8-10] because of the high group velocity dispersion (GVD) in the slow-light optical spectrum range. To broaden the operating optical bandwidth of PCW modulators, lattice shifted PCWs can be employed, where the spatial shift of certain holes can modify the structure to provide low-dispersion slow light [11-15].

In this letter we report a symmetric MZI modulator based on band-engineered slot PCW refilled with EO polymer, SEO125 from Soluxra, LLC. SEO125 exhibits exceptional combination of large EO activity, low optical loss, and good temporal stability. Its $r_{33}$ value of poled thin films is around 125pm/V at the wavelength of 1310 nm, which is measured by the Teng–Man reflection technique. The design and synthesis of SEO125 encompasses recent development of highly efficient nonlinear optical chromophores with a few key molecular and material parameters, including large β values, good near-infrared transparency, excellent chemical- and photo-stability, and improved processability in polymers [16]. Using a band-engineered EO polymer refilled slot PCW with $S_w$=320nm, we demonstrate a slow-light enhanced effective in-device $r_{33}$ of 1190pm/V over 8nm optical spectrum range. Excluding the slow-light effect, we estimate in-device material' $r_{33}$ of 89pm/V for SEO125 in the slot that show 51% improvement compared to the results (59pm/V) in [7].

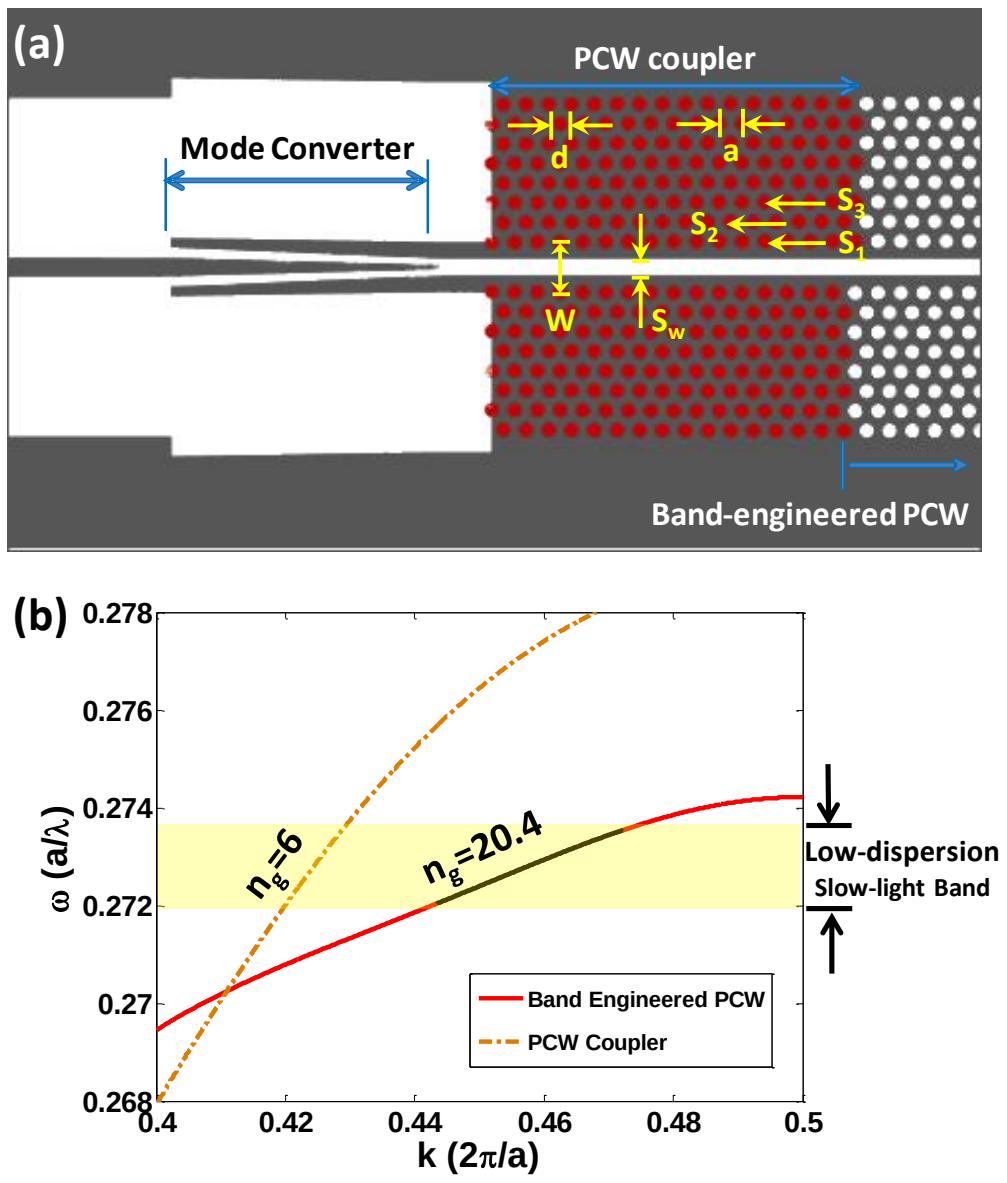

Fig. 1. (a) Layout of the PCW coupler (mode converter + PCW coupler). The black area corresponds to un-etched silicon. (b) Band diagram of the engineered slow-light PCW and the PCW coupler.

A schematic of the device on silicon on insulator (SOI) (Si thickness=250nm, oxide thickness=3μm) is shown in Fig. 1 (a). The input and output strip waveguides are connected to the device using a strip-to-slot waveguide mode converter. PCW couplers consisting of a fast-light section [17] connect the mode converters to a 300μm-long slow-light PCW section. The slow-light PCW section is band-engineered by lateral shifting of the first three rows on the two sides of the slot [indicated by $s_1$, $s_2$, $s_3$ in Fig. 1 (a)] [12] and by varying the center-to-center distance between two rows adjacent to the slot [W in Fig. 1 (a)]. Multi-mode interference (MMI) couplers are used for beam splitting/combining [18]. Sub-wavelength grating (SWG) are designed to couple light into and out of the silicon strip [19].

For lattice constant, a=425nm, it is found that with a hole diameter d=300nm, $s_1$=0, $s_2$=-85nm, $s_3$=85nm, $S_w$=320nm, and W=1.54(√3)a, we can achieve an average group index ($n_g$=c/$v_g$) of 20.4 (±10%) over 8.2nm optical bandwidth. The PCW coupler [a=425nm, d=300nm, $s_1$=0, $s_2$=0, $s_3$=0, $S_w$=320nm, W=1.45(√3)a] consists of 16 periods and is designed for low $n_g$=6 over the same wavelength range. The band diagrams of the slow-light and fast-light PCWs are shown in Fig. 1 (b).

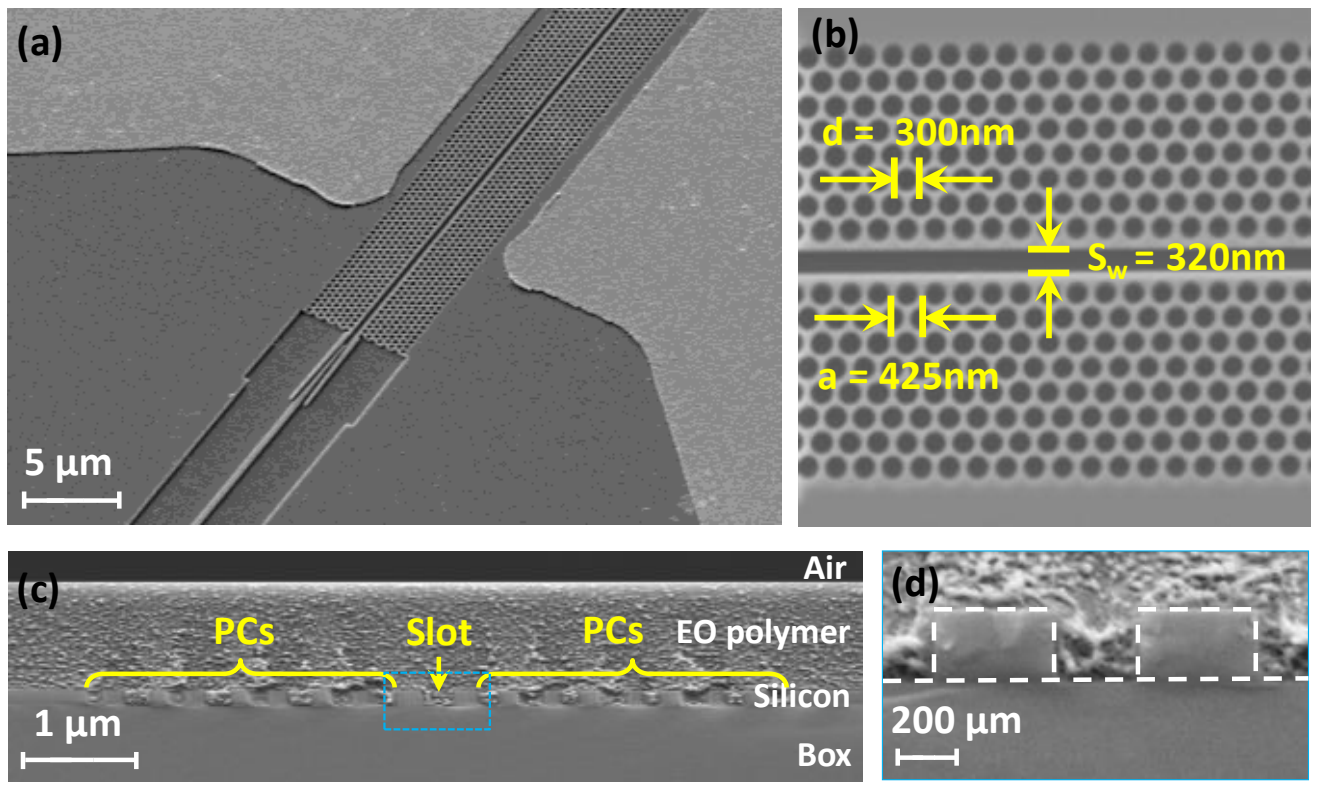

Fig. 2. SEM images of the fabricated device. (a) Tiled view of a local area of silicon slot PCW modulator. (b) Top view of slot PCW area. (c) Cross-sectional view of the EO polymer refilled silicon slot PCW. PCs: photonic crystals. (d) Zoom-in image of the dashed square area in (c).

The fabrication procedure starts with an SOI wafer with 250nm-thick top silicon. All the photonic circuitries are fabricated using electron-beam lithography and reactive ion etching (RIE) in a single patterning/etching step, while the gold electrodes are patterned by photolithography and lift-off process, as shown in Fig. 2 (a) and (b). The EO polymer, SEO125, is infiltrated into the slot PCW by spincoating. The silicon PCW regions including holes and the slot are fully covered by EO polymer, as shown in the SEM image in Fig. 2 (c) and (d). A microscope image of the fabricated MZI is shown in Fig. 3 (a). Next, the sample is poled by an electric field of 100V/µm in a push-pull configuration at the glass transition temperature ($T_g$=145 ℃) of the EO polymer. The leakage current as well as the hot plate temperature is monitored and shown in Fig. 3 (b). It can be seen that the maximum leakage current remains below 0.659nA, corresponding to leakage current density of 8.79A/$m^2$ [=0.659nA/(300um*250nm)]. For comparison, the typical leakage current density of the EO polymer is 1-10A/$m^2$ in a thin film configuration. This poling result is repeatable and shows that the 320nm-wide slot dramatically reduces the leakage current that is known to be detrimental to the poling efficiency [20].

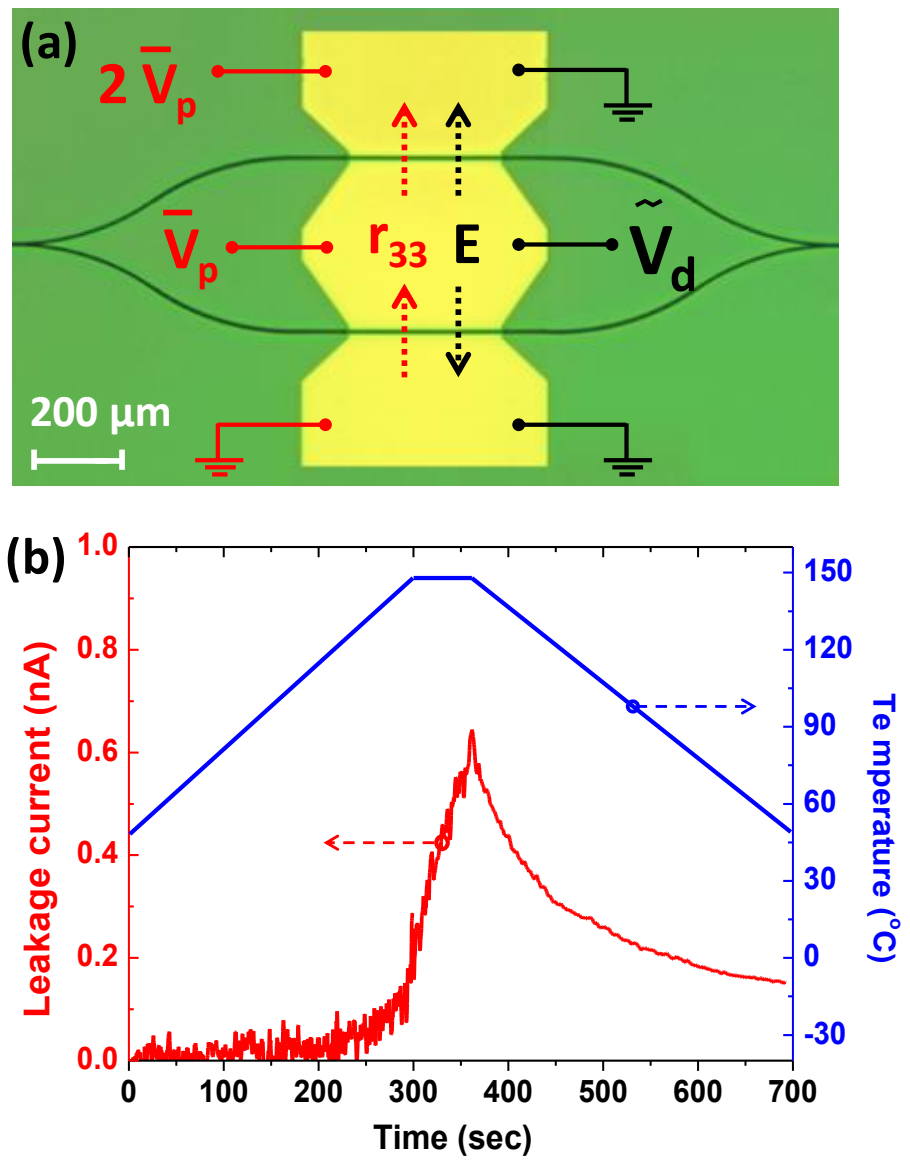

Fig. 3. (a) Top view of fabricated slot PCW MZI modulator. The red colored circuit connection indicates the push-pull poling configuration and induced $r_{33}$ direction, and the black colored circuit connection indicates the modulation configuration. $V_p$: poling voltage, $V_d$: diving voltage. (b) The temperature-dependent leakage current in the EO polymer poling process.

For modulation test, TE-polarized light from a tunable laser source (1550nm, 2.5mW) is coupled into and out of the device through grating couplers. The total optical insertion loss is 20dB, including the 6.5dB/facet coupling loss from grating couplers. RF signals are applied to the electrodes as shown in Fig. 3 (a). The modulator is biased at the 3dB point and driven by a 100KHz triangular RF wave with a peak-to-peak voltage of 1.4V. The modulated output optical signal is sent to a photodetector and then displayed on a digital oscilloscope. The modulation frequency is within the bandwidth of the photodetector and the oscilloscope. From the output optical waveform measured by the digital oscilloscope, over-modulation is observed. The $V_\pi$ of the modulator is measured to be 0.973V from the transfer function of the over-modulated optical signal and the input RF signal on the oscilloscope, by finding the difference between the applied voltage at which the optical output is at a maximum and the voltage at which the optical output is at the following minimum [1, 3, 7, 20, 21]. The effective in-device $r_{33}$ is then calculated to be

$$r_{33-\text{effective}} = \frac{\lambda S_w}{n^3 V_\pi \sigma L} = 1190\text{pm/V} \qquad (1)$$

where, λ=1.55 µm, $S_w$=320nm, n=1.63, L=300µm, σ=0.33 (confinement factor in the slot) calculated by simulation. This extraordinarily high $r_{33}$ value confirms the combined enhancing effects of slow light and an improved poling efficiency. This band-engineered 320nm slot PCW modulator also achieves very high modulation efficiency with $V_\pi$×L=0.973V ×300µm=0.292V×mm.

We also estimate the actual in-device $r_{33}$ excluding the slow-light effect using [11]

$$L = \frac{\lambda}{2\sigma n_g}\left(\frac{n}{\Delta n}\right) \qquad (2)$$

where, $\Delta n = n^3 r_{33} V_\pi / (2S_w)$. The estimated in-device $r_{33}$ is 89pm/V that is significantly larger than our previous work in [7] and is the highest poling efficiency demonstrated in a slot waveguide to the best of our knowledge. Considering the $r_{33}$ dispersion from the two-level model approximation [22], this value also represents nearly 100% poling efficiency that has been obtained in poled thin films of SEO125.

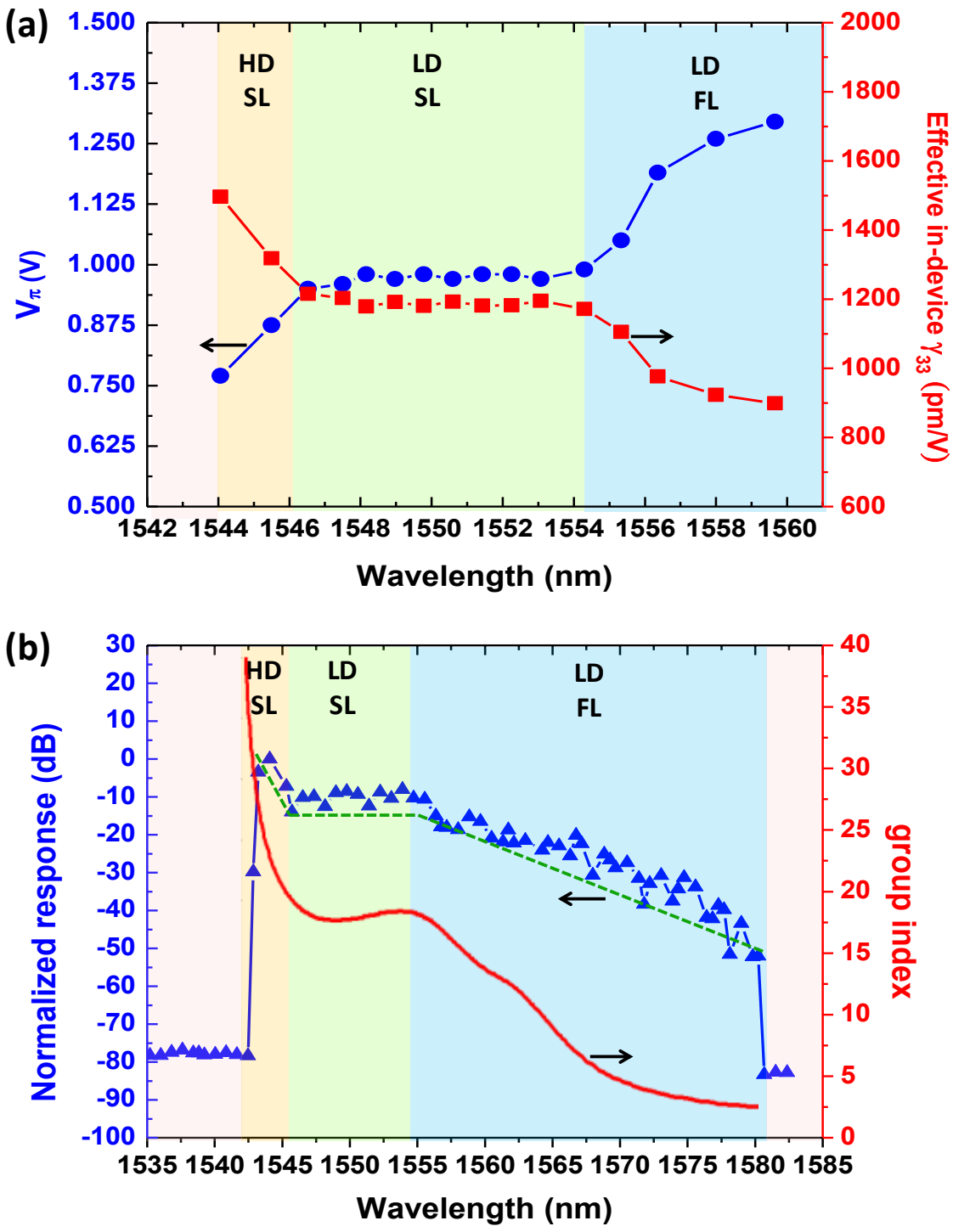

Fig. 4. (a) Measured $V_\pi$ and corresponding calculated effective in-device $r_{33}$ v.s. wavelength (at 100KHz). HD SL: high-dispersion slow-light; LD SL: low-dispersion slow-light; LD FL: low-dispersion fast-light. (b) Normalized device response v.s. wavelength (at 100KHz). The green dashed line indicates the trend of the response change over different wavelength. The simulated $n_g$ v.s. wavelength is also overlaid.

To demonstrate the wide optical spectrum range, the optical wavelength is tuned from 1544nm to 1560nm while all other testing conditions are fixed. The $V_\pi$ measured at different wavelength, as well as the corresponding calculated effective in-device $r_{33}$, is plotted in Fig. 4 (a). It can be seen that the $V_\pi$ is nearly constant, which is 0.97±0.02V, over optical spectrum range of 8nm (low-dispersion slow-light region: from 1546.5nm to 1554.5nm), corresponding to the effective in-device $r_{33}$ of 1190pm/V and $V_\pi \times L$ of 0.291 ± 0.006V×mm. We note that this $V_\pi \times L$ value is relative to a push-pull configuration. Relative to a single-arm modulator where the effective length of the MZI is the length of both arms together, $V_\pi \times (2L)$ = 0.582 ±0.012V×mm is still a record low value.

Furthermore, a small signal modulation test is done at $V_{pp}$<1V over a range of wavelength from 1535nm to 1582nm, while all other testing conditions remain the same. The modulated optical signal is converted to electrical signal by a photodetector and then measured by a microwave spectrum analyzer. The wavelength dependence of the normalized modulated optical signal is plotted in Fig. 4 (b). It can be seen that the defect-guided mode of slot PCW occurs from 1543nm to 1580nm. The maximum response occurs at the high-dispersion slow-light region (wavelength from 1543nm to 1546.5nm), because of the largest slow-light enhancement (largest $n_g$) in this region. The response is almost flat in the low-dispersion slow-light region (wavelength from 1546.5nm to 1554.5nm), because the slot PCW is band-engineered to have a nearly constant $n_g$ in this wavelength range. As the optical signal is tuned to longer wavelength (low-dispersion fast-light region: from 1554.5nm to1580nm), the device response becomes smaller due to decreasing $n_g$.

In summary, we design, fabricate and characterize a band-engineered EO polymer refilled silicon slot PCW MZI modulator. The half-wave switching-voltage is measured to be $V_\pi$=0.97±0.02V over optical spectrum range of 8nm, corresponding to the slow-light enhanced effective in-device $r_{33}$ of 1190pm/V and $V_\pi \times L$ of 0.291±0.006V×mm. Excluding the slow-light effect, we estimate the EO polymer is poled with a record-high EO activity of 89pm/V in the slot at the wavelength of 1.55μm. In our future work, the optical loss of our modulator can be further reduced, such as by the design of low-loss PCW couplers [23] and improved coupling and packaging method [24]. The photochemical stability, a common issue for polymer based modulators, is expected to be improved by hermetically sealing of EO polymer in a robust packaging [25, 26]. Poled thin films of SEO125 have shown good temporal stability due to its relatively high $T_g$=145℃, and after the poling its EO coefficients were essentially unchanged under ambient conditions. While SEO125 is a newly developed material and its complete characterization in terms of performance and photo-stability is an ongoing effort, EO polymers with similar compositions have been demonstrated to have potential long-term stability by removing oxygen in the packaging of devices [27].

The authors would like to acknowledge the Air Force Research Laboratory (AFRL) for supporting this work under the Small Business Technology Transfer Research (STTR) program (Grant No. FA8650-12-M-5131) monitored by Dr. Robert Nelson and Dr. Charles Lee. Amir Hosseini and Xingyu Zhang contributed to this work equally.

**Reference**

1. Y. Shi, C. Zhang, H. Zhang, J. H. Bechtel, L. R. Dalton, B. H. Robinson, and W. H. Steier, "Low (sub-1-volt) halfwave voltage polymeric electro-optic modulators achieved by controlling chromophore shape," Science **288**, 119-122 (2000).
2. D. Chen, H. R. Fetterman, A. Chen, W. H. Steier, L. R. Dalton, W. Wang, and Y. Shi, "Demonstration of 110 GHz electro-optic polymer modulators," Applied Physics Letters **70**, 3335-3337 (1997).
3. X. Zhang, A. Hosseini, X. Lin, H. Subbaraman, and R. T. Chen, "Polymer-based Hybrid Integrated Photonic Devices for Silicon On-chip Modulation and Board-level Optical Interconnects," IEEE Journal of Selected Topics in Quantum Electronics **16**, 3401115-3401115 (2013).
4. R. Ding, T. Baehr-Jones, W.-J. Kim, A. Spott, M. Fournier, J.-M. Fedeli, S. Huang, J. Luo, A. K.-Y. Jen, and L. Dalton, "Sub-volt silicon-organic electro-optic modulator with 500 MHz bandwidth," Journal of Lightwave Technology **29**, 1112-1117 (2011).

5. M. Gould, T. Baehr-Jones, R. Ding, S. Huang, J. Luo, A. K.-Y. Jen, J.-M. Fedeli, M. Fournier, and M. Hochberg, "Silicon-polymer hybrid slot waveguide ring-resonator modulator," Optics express **19**, 3952-3961 (2011).
6. J. H. Wülbern, S. Prorok, J. Hampe, A. Petrov, M. Eich, J. Luo, A. K.-Y. Jen, M. Jenett, and A. Jacob, "40 GHz electro-optic modulation in hybrid silicon–organic slotted photonic crystal waveguides," Optics letters **35**, 2753-2755 (2010).
7. X. Wang, C.-Y. Lin, S. Chakravarty, J. Luo, A. K.-Y. Jen, and R. T. Chen, "Effective in-device r< sub> 33</sub> of 735 pm/V on electro-optic polymer infiltrated silicon photonic crystal slot waveguides," Optics letters **36**, 882-884 (2011).
8. H. C. Nguyen, Y. Sakai, M. Shinkawa, N. Ishikura, and T. Baba, "10 Gb/s operation of photonic crystal silicon optical modulators," Optics Express **19**, 13000-13007 (2011).
9. J. H. Wulbern, J. Hampe, A. Petrov, M. Eich, J. Luo, A. K.-Y. Jen, A. Di Falco, T. F. Krauss, and J. Bruns, "Electro-optic modulation in slotted resonant photonic crystal heterostructures," Applied Physics Letters **94**, 241107-241107-241103 (2009).
10. H. C. Nguyen, Y. Sakai, M. Shinkawa, N. Ishikura, and T. Baba, "Photonic crystal silicon optical modulators: carrier-injection and depletion at 10 Gb/s," Quantum Electronics, IEEE Journal of **48**, 210-220 (2012).
11. A. Hosseini, X. Xu, H. Subbaraman, C.-Y. Lin, S. Rahimi, and R. T. Chen, "Large optical spectral range dispersion engineered silicon-based photonic crystal waveguide modulator," Opt. Express20 (11), 12318-12325 (2012).
12. S. Rahimi, A. Hosseini, X. Xu, H. Subbaraman, and R. T. Chen, "Group-index independent coupling to band engineered SOI photonic crystal waveguide with large slow-down factor," Opt. Express19 (22), 21832-21841 (2011).
13. Y. Hamachi, S. Kubo, and T. Baba, "Slow light with low dispersion and nonlinear enhancement in a lattice-shifted photonic crystal waveguide," Optics letters **34**, 1072-1074 (2009).
14. S. Schulz, L. O'Faolain, D. Beggs, T. White, A. Melloni, and T. Krauss, "Dispersion engineered slow light in photonic crystals: a comparison," Journal of Optics **12**, 104004 (2010).
15. A. Y. Petrov, and M. Eich, "Zero dispersion at small group velocities in photonic crystal waveguides," Applied Physics Letters **85**, 4866-4868 (2004).
16. J. Luo, X.-H. Zhou, and A. K.-Y. Jen, "Rational molecular design and supramolecular assembly of highly efficient organic electro-optic materials," Journal of Materials Chemistry **19**, 7410-7424 (2009).
17. A. Hosseini, X. Xu, D. N. Kwong, H. Subbaraman, W. Jiang, and R. T. Chen, "On the role of evanescent modes and group index tapering in slow light photonic crystal waveguide coupling efficiency," Applied Physics Letters **98**, 031107-031107-031103 (2011).
18. A. Hosseini, D. Kwong, C.-Y. Lin, B. S. Lee, and R. T. Chen, "Output Formulation for Symmetrically Excited One-to-< formula formulatype=," Selected Topics in Quantum Electronics, IEEE Journal of **16**, 61-69 (2010).
19. X. Xu, H. Subbaraman, J. Covey, D. Kwong, A. Hosseini, and R. T. Chen, "Complementary metal–oxide–semiconductor compatible high efficiency subwavelength grating couplers for silicon integrated photonics," Applied Physics Letters **101**, 031109-031109-031104 (2012).
20. X. Zhang, B. Lee, C.-y. Lin, A. X. Wang, A. Hosseini, and R. T. Chen, "Highly Linear Broadband Optical Modulator Based on Electro-Optic Polymer," Photonics Journal, IEEE **4**, 2214-2228 (2012).
21. C.-Y. Lin, X. Wang, S. Chakravarty, B. S. Lee, W. Lai, J. Luo, A. K.-Y. Jen, and R. T. Chen, "Electro-optic polymer infiltrated silicon photonic crystal slot waveguide modulator with 23 dB slow light enhancement," Applied Physics Letters **97**, 093304-093304-093303 (2010).
22. C. Greenlee, A. Guilmo, A. Opadeyi, R. Himmelhuber, R. A. Norwood, M. Fallahi, J. Luo, S. Huang, X.-H. Zhou, and A. K.-Y. Jen, "Mach–Zehnder interferometry method for decoupling electro-optic and piezoelectric effects in poled polymer films," Applied Physics Letters **97**, 041109-041109-041103 (2010).
23. R. Palmer, L. Alloatti, D. Korn, W. Heni, P. C. Schindler, J. Bolten, M. Karl, M. Waldow, T. Wahlbrink, W. Freude, C. Koos, and J. Leuthold, "Low-Loss Silicon Strip-to-Slot Mode Converters," Ieee Photonics J **5** (2013).
24. B. Snyder, and P. O'Brien, "Planar fiber packaging method for silicon photonic integrated circuits," in *Optical Fiber Communication Conference and Exposition (OFC/NFOEC), 2012 and the National Fiber Optic Engineers Conference*(IEEE2012), pp. 1-3.
25. R. Dinu, D. Jin, G. M. Yu, B. Q. Chen, D. Y. Huang, H. Chen, A. Barklund, E. Miller, C. L. Wei, and J. Vemagiri, "Environmental Stress Testing of Electro-Optic Polymer Modulators," Journal of Lightwave Technology **27**, 1527-1532 (2009).
26. D. Jin, H. Chen, A. Barklund, J. Mallari, G. Yu, E. Miller, and R. Dinu, "EO polymer modulators reliability study," in *OPTO*(International Society for Optics and Photonics2010), pp. 75990H-75990H-75998.
27. S. Takahashi, B. Bhola, A. Yick, W. H. Steier, J. Luo, A. K.-Y. Jen, D. Jin, and R. Dinu, "Photo-Stability Measurement of Electro-Optic Polymer Waveguides With High Intensity at 1550-nm Wavelength," Journal of Lightwave Technology **27**, 1045-1050 (2009).